\documentclass[twocolumn,times]{aastex631}

\newcommand{\td}{{\sc technicolor dawn}}

\newcommand{\NHI}{$N_{\mathrm{HI}}\,$}

\usepackage{breqn}
\usepackage[normalem]{ulem}

\shorttitle{Galaxy PDLAs and Reionization Topology in \td}
\shortauthors{Steen et al.}
\graphicspath{{./}{figures/}}
\usepackage{lipsum}
\usepackage{soul}
\begin{document}

\title{Galaxy Proximate Damped Lyman-Alpha Systems and HI Reionization Topology in TECHNICOLOR DAWN}

\author[0009-0002-0435-5055]{Maya Steen}
\affiliation{New Mexico State University \\
MSC 4500, PO BOX 30001 \\
Las Cruces, NM 88003}

\author[0000-0002-0496-1656]{Kristian Finlator}
\affiliation{New Mexico State University \\
MSC 4500, PO BOX 30001 \\
Las Cruces, NM 88003}
\affiliation{Cosmic Dawn Center (DAWN) \\
Niels Bohr Institute, University of Copenhagen / DTU-Space \\ 
Technical University of Denmark}

\author[0000-0002-0761-1985]{Samir Ku\v{s}mi\'{c}}
\affiliation{New Mexico State University \\
MSC 4500, PO BOX 30001 \\
Las Cruces, NM 88003}

\author[0009-0004-8503-0483]{Ezra Huscher}
\affiliation{New Mexico State University \\
MSC 4500, PO BOX 30001 \\
Las Cruces, NM 88003}

\begin{abstract}

Recent observations from the James Webb Space Telescope (JWST) have revealed proximate damped Lyman-$\alpha$ systems (PDLAs) in the foreground of high redshift galaxies ($z > 5$), which have been interpreted as neutral circumgalactic media (CGM). The ionization state of the CGM, potentially inferred from DLA strength, may serve as a probe to trace the progress of reionization, similarly to the ionization state of the intergalactic medium (IGM). To determine if this method has merit, we use the cosmological hydrodynamical simulation TECHNICOLOR DAWN to study simulated gas halos at redshifts $z = 10, 8, 6, \text{ and } 5.5$. We investigate the reionization topology to determine whether the CGM and IGM have similar ionization histories, and we study the relation between column density of neutral hydrogen (observationally measured by DLA strength), neutral fraction, and gas mass fraction of the foreground gas to determine whether PDLAs can be used to trace the progress of reionization. We find an inside-out-middle reionization topology, where the CGM reionizes after the IGM and remains partially neutral at $ z= 5.5$. The foreground column density of neutral hydrogen depends mostly on halo mass, with a weak dependence on neutral fraction or redshift. Therefore, provided precise estimates of halo or stellar mass, PDLAs may be used to trace the progress of reionization particularly at high redshifts.

\end{abstract}

 \section{Introduction} \label{sec:intro}

Recent analysis of data from the James Webb Space Telescope (JWST) 
by \cite{heintz_strong_2024, heintz_jwst-primal_2025} 
has led to speculation that proximate damped Lyman-$\alpha$ absorbers (PDLAs) may be a new way to trace the progress of reionization. 
The presence of PDLAs at high redshifts suggests a massive, neutral reservoir around these galaxies, i.e. a large, neutral circumgalactic medium (CGM). 

DLAs are well known in the foreground of quasars (QSOs), called proximate DLAs (PDLAs) when nearby to the quasar in relative velocity space. These PDLAs have been used to study the quasar host galaxy’s environment or ejecta \citep{ellison_nature_2010}. We investigate a potential new approach by using DLAs proximate to galaxies to study the CGM of high redshift galaxies. These galaxy PDLAs have only been observed for high redshift galaxies, before the completion of global reionization. 


Reionization of neutral hydrogen marks the last major phase change of the universe. This is the process in which early luminous sources produced ionizing radiation, which gradually changed the ionization state of the surrounding gas from neutral to ionized. Various observations point to early galaxies as being the main source of ionizing radiation contributing to cosmic reionization  (e.g. \cite{ dodorico_metals_2013, bouwens_reionization_2015, begley_evolution_2025}). The process of reionization is known to be highly inhomogeneous due to density perturbations in the early universe that have propagated as the universe has expanded, the peaks of which seed the galaxies observe today \citep{miralda-escude_reionization_1998, miralda-escude_reionization_2000, dawoodbhoy_cosmic_2023}. While progress has been made in studying reionization, the topology and timing are still relatively unconstrained. New simulations and observations are needed to more tightly constrain both. 

The topology of reionization is mainly studied through simulations. It is expected that reionization followed either an inside-out or outside-in topology, driven by ionizing radiation escaping from early galaxies.  

Inside-out reionization begins with early bright sources, such as galaxies and quasars, which emit ionizing flux into their immediate surroundings.  \citep{iliev_simulating_2006, lee_topology_2008}. This forms HII ``bubbles'' in these high-density regions that grow as flux is able to penetrate further into the surrounding gas \citep{iliev_simulating_2006}. Because galaxies are so clustered, these HII bubbles merge quickly \citep{lee_topology_2008}. 
The result is that high-density, galaxy-hosting regions are reionized first, and low density regions far from ionizing sources are reionized later  \citep{iliev_simulating_2006, lee_topology_2008}, with some dense neutral clumps remaining until late times as they can be self-shielded \citep{lee_topology_2008}. 

Outside-in reionization topology, on the other hand, results in low density regions being reionized first, with high density regions being reionized later \citep{miralda-escude_reionization_2000, doughty_evolution_2019}. Similarly to inside-out reionization, this begins with ionizing flux emitted from early bright sources. However, instead of reionizing the immediate surroundings, the ionizing flux “punches holes” in the dense media and escapes to less dense regions \citep{doughty_evolution_2019}. This creates ionized pockets within dense media, but these are small and leave dense regions mostly neutral on average \citep{miralda-escude_reionization_2000}. The escaping ionizing flux is able to quickly ionize the IGM, as low density regions have a lower recombination rate working against ionization, as well as simply having fewer atoms per volume to be ionized \citep{doughty_evolution_2019}. This process creates large cosmological HII regions in the IGM, which grow in the direction of lowest density as the ionization fronts propagate \citep{miralda-escude_reionization_2000}. These HII regions eventually overlap, and the ionization fronts work back in towards regions of higher density. Again, only a few high density neutral clouds can remain due to self-shielding \citep{miralda-escude_reionization_2000}. 

\cite{finlator_late_2009} finds a combination of these cases, referred to as “inside-out-middle” (IOM) topology. In this situation, reionization first proceeds in an inside out manner, where dense regions immediately surrounding ionization sources are reionized first. Ionizing flux then leaks directly into the least dense regions, due to the low recombination rates there. Reionization then works back in, following a more outside-in reionization topology, and reionizes the intermediate density regions (filaments) last, due to high recombination rates and low emissivity in these regions \citep{finlator_late_2009}.



Unlike studying the topology of reionization, the timing of reionization can be constrained using various observational methods.nThe Lyman-$\alpha$ forest (LAF) is a series of absorption features in the foreground of quasars
, where bluer light than Ly-$\alpha$ is redshifted into resonance and absorbed by neutral hydrogen in foreground gas clouds \citep{fan_constraining_2006, becker_evidence_2015, bosman_hydrogen_2022}. When the neutral fraction of these clouds reaches a moderate level, these features become saturated and blend together into the Gunn Peterson (GP) trough \citep{gunn_density_1965}. 
Strong variations in the size and distribution of GP troughs across different sightlines as well as scatter in the LAF imply that neutral hydrogen persists at $z<6$ and that the ultraviolet background is inhomogeneous at this time, suggestive of a late ($z \approx 5$) end to reionization \citep{becker_evidence_2015, bosman_hydrogen_2022}. Some quasar spectra also show damping wings to the red side of the Lyman-$\alpha$ line, called the Gunn Peterson damping wing \citep{miralda-escude_reionization_1998, schroeder_evidence_2013}. The width of this wing traces the neutral fraction of the intergalactic medium (IGM) \citep{miralda-escude_reionization_1998}. The optical depth in the damping wing is orders of magnitude lower than in the core of the line, therefore to detect a damping wing the neutral fraction must be significant, unlike the GP trough and LAF which saturate at low neutral fractions \citep{schroeder_evidence_2013}.

Metal lines in quasar sightlines can also constrain the timing of reionization, as these lines can be studied even in the case of a fully saturated GP trough \citep{oppenheimer_tracing_2009}. 
OI is of interest because it has a similar ionization energy to neutral hydrogen and is therefore sensitive to the same radiation that drives HI reionization. Unlike the assumed picture of metallicity increasing over time with stellar enrichment, the number density OI of absorbers decreases over time, implying this gas transitions to a higher ionization state at lower redshifts due to the strengthening of the ultraviolet background \citep{becker_evolution_2019}.
Simulated OI absorbers in \cite{doughty_evolution_2019} further support that this evolution in detected OI absorbers is due to an evolving UVB, particularly affecting the weak absorbers. Variations in both strong and weak absorbers observed across various sightlines at $z>5$ suggests that reionization may still be occurring at this time \citep{becker_high-redshift_2009}. Although these methods work well to constrain the end of reionization, Quasars and GRBs are rare at $z>7$, limiting the use of methods that rely on these sources. 

Lyman-$\alpha$ Emitters (LAEs) are also useful at probing reionization, as the emitted Ly-$\alpha$ is sensitive to the neutral hydrogen fraction of the nearby IGM \citep{inoue_silverrush_2018}. Evolution in the LAE luminosity function with redshift can therefore show changes in the IGM and suggests an increasing neutral fraction when $z>6$ as a consequence of reionization \citep{inoue_silverrush_2018}, but requires assumptions about ``intrinsic'' underlying Lyman-$\alpha$ EW distribution. This may also be contaminated with effects from galaxy evolution \citep{konno_silverrush_2018}. 

Finally, new cosmic microwave background measurements constrain the Thomson optical depth from the last scattering surface from ionized material using their new ``low multipole likelihood’’ method for polarization maps \citep{planck_collaboration_planck_2020}. These result in a reionization midpoint of $z \approx 7.7$, which can be compared to other methods of measuring the timing of reionization \citep{planck_collaboration_planck_2020}. However, interpreting CMB data requires an assumed model for reionization.




In this work, we investigate a new method for tracing the progress of reionization: galaxy PDLAs as observed by JWST. 
Here, the idea is to measure the IGM's ionization state indirectly by instead tracing that of the high-redshift CGM. 
However, to confirm the merit of this method, we must answer the following questions. First, in order to use the CGM to probe reionization progress, we must determine if the CGM and IGM have similar reionization histories according to the topology of reionization. 
Second, the excess absorption implied by galaxy PDLAs could reflect high neutral fractions (local or global), high halo gas fractions, or some combination of both. For PDLAs to trace reionization, a clear correlation between PDLA strength (or column density of neutral hydrogen) and global neutral fraction should be found.

We use results from the \td\, simulation to investigate these questions, and address if these DLAs detected proximate to high redshift galaxies may provide a new way to study the progress of reionization. In Section \ref{sec:sim}, we discuss the details of \td, the simulation used to complete this analysis. In Section \ref{sec:results} we will first discuss the topology of reionization, which is traced using the neutral fraction for regions of different overdensities as a function of redshift. We will then move to a halo-by-halo analysis of the CGM, where we will study relations between various CGM properties including the column density of neutral hydrogen. This will indicate if high column densities are a product of largely neutral CGM or simply from more gas mass in the CGM. This analysis will also determine which characteristics of the CGM along with inferred DLA parameters may be used to trace reionization observationally. We discuss the results from this analysis and compare to existing literature in Section \ref{sec:discussion}. We summarize and discuss future work relevant to our findings in Section \ref{sec: summary}.  


\section{Simulation} \label{sec:sim}

\begin{figure}
\centering{\includegraphics[width=0.998\columnwidth]{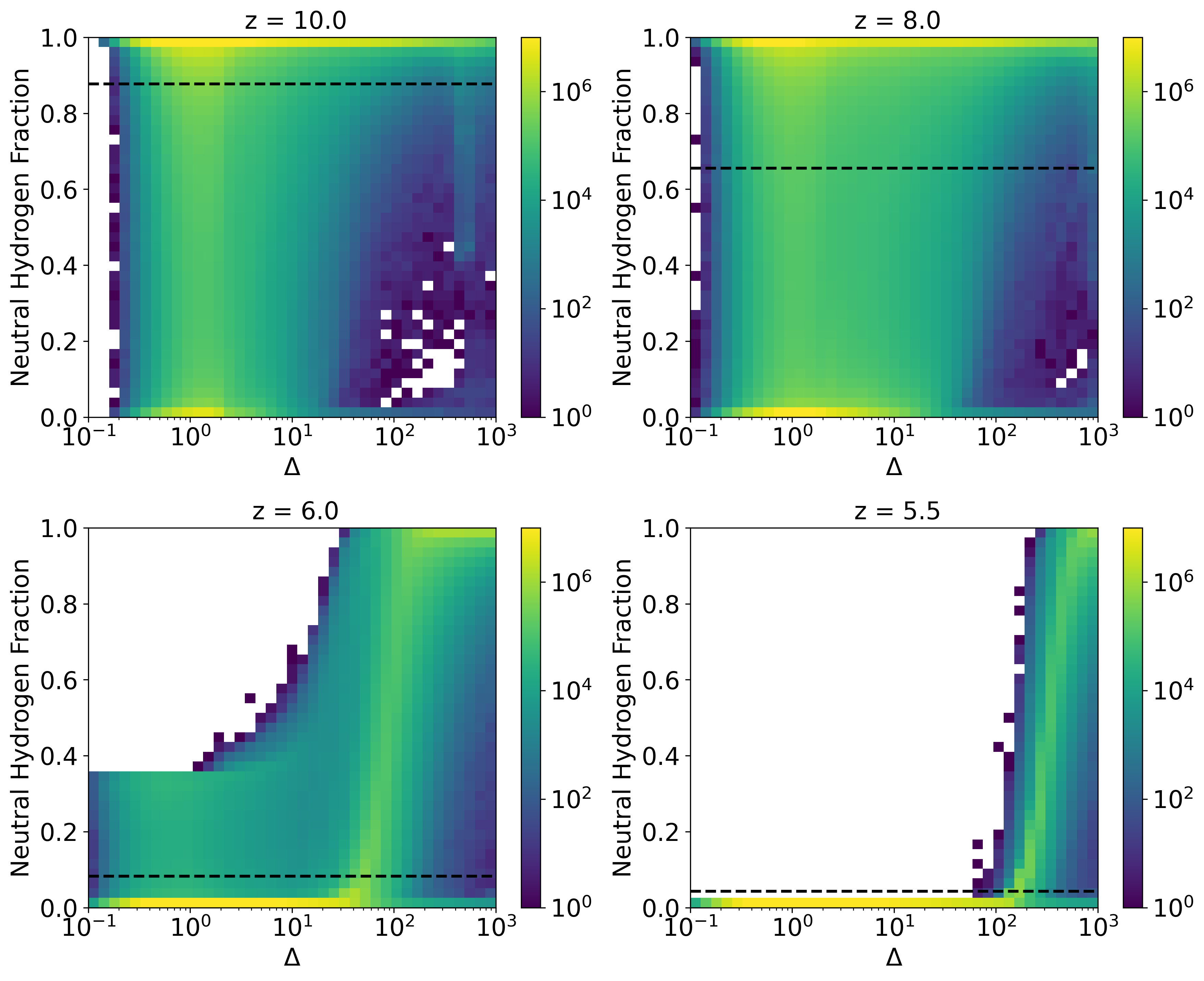}}
\caption{Two-dimensional histogram showing neutral hydrogen fraction versus overdensity for $z=10,\, 8,\, 6,$ and $5.5$. The color bar represents the number of particles in that bin, and white bins contain no particles. The black horizontal line shows the volume-averaged neutral fraction at each respective redshift. Reionization broadly progresses from lower to higher overdensity regions (outside-in), although a small number of intermediate density regions are ionized even when the Universe remains largely neutral ($z \approx 10$). 
\label{fig:ezra_plot}}
\end{figure}

\td \, is a hydrodynamical radiation cosmological simulation with 16.5 cMpc/h side lengths and initialized with $2 \times 704^3$ particles including gas and dark matter \citep{finlator_reionization_2018}. The metagalactic ionizing background is modeled on the fly via the moment method, discretized spatially into $88^3$ cubical voxels and spectrally into 32 frequency bins spanning 1--10 Rydbergs (Ryd) \citep{finlator_reionization_2018}.
The cosmological model is $\mathrm{\Lambda}$CDM. This simulation uses the Planck cosmological parameters, which are $h = 0.6774$, $\Omega_{\mathrm{\Lambda}} = 0.6911$, $\Omega_m = 0.3089$, $\Omega_b = 0.0486$, and $X_H = 0.751$ \citep{planck_collaboration_planck_2016}.

Gas particles are treated through a smoothed particle hydrodynamics (SPH) prescription based on the SPH framework in Gadget-3 \citep{springel_cosmological_2005}. Particles are allowed to cool radiatively from collisional excitation of both H and He, and gas particles with nonzero metallicity can also cool via collisional excitation of metals \citep{finlator_reionization_2018}. 
We use an updated run of \td,  featuring updated feedback and physical modeling as well as improved dynamic range (detailed in \cite{finlator_faint_2020} section 2 and \cite{huscher_impact_2024} section 2), to test which of the the previously suggested reionization models are present in this latest simulation run: the outside-in reionization model from \cite{finlator_reionization_2018} or the inside-out-middle model present in \cite{finlator_late_2009}.

\section{Analysis and Results} \label{sec:results}


We investigate how reionization progresses through regions of different densities to compare the reionization histories of the IGM and the CGM. We also study the CGM of individual halos to determine how the column density of neutral hydrogen, as implied by the observed PDLAs in galaxy spectra, is related to other halo properties, such as gas fraction and neutral fraction of the halo gas. 

If the CGM and IGM have similar reionization histories, the CGM can be used in a similar fashion as the IGM as a tracer of reionization. Additionally, if the column density of neutral hydrogen is correlated to the global neutral fraction, then column density can be used to probe the progress of reionization. Therefore, the combination of these two results will allow us to determine if galaxy PDLAs can be used as a new observational method to measure the progress of reionization. 

\td\, is well suited for this problem, as it has sufficient resolution to probe IGM and CGM gas, as well as containing a sufficiently large volume to grow a robust sample of galaxies. We conduct our analysis in two different methods, particle-by-particle and halo-by-halo. 


In Section \ref{subsec:top} we investigate the topology of reionization in the newest \td\, simulation run. Here, we approach the problem on a particle-by-particle basis, allowing us to track the progress of reionization across voids, the IGM, and the CGM. There is a general progression of reionization from least dense to most dense regions.

In Section \ref{subsec:cgm} we now study the simulation halo-by-halo to study properties of the CGM and galaxies. Here, we investigate how column density of neutral hydrogen may be related to other halo properties, such as the mass, neutral fraction, or gas mass fraction and global properties such as the volume averaged neutral fraction. This allows us to determine if column density of neutral hydrogen, derived from observed DLAs proximate to galaxies, can be used to help constrain the progress of reionization.

\subsection{Topology of Reionization}\label{subsec:top}

\begin{figure}
\centering{\includegraphics[width=0.998\columnwidth]{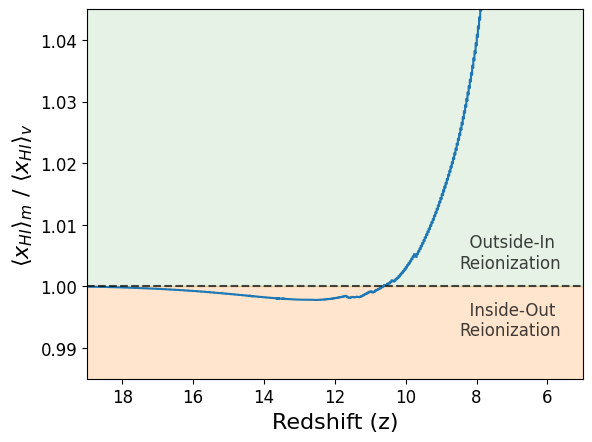}}
\caption{Evolution of the ratio of the mass-averaged to volume-averaged neutral hydrogen fraction, $\langle x_{HI}\rangle_m \, / \, \langle x_{HI}\rangle_v$, as a function of redshift. Ratio values of less than one (orange shaded region) indicate an inside-out reionization progression, while ratio values of greater than one (green shaded region) indicate an outside-in progression. In this representation, we see a brief, early inside-out reionization followed by a longer and more dominant outside-in phase. This combination of reionization topologies can also be called inside-out-middle (IOM) reionization, similar to that in \cite{finlator_late_2009}. 
\label{fig:topology}}
\end{figure}

To investigate the topology of reionization, we first compute the overdensity $\Delta$ of each particle by following Equations \ref{eqn:avg_den} and \ref{eqn:delta}, where $\langle \rho \rangle$ is the volume weighted average density, $V_i$ is the volume of each particle, and $\rho_i$ the density for each gas particle. 


\begin{equation} \label{eqn:avg_den} 
    \langle \rho \rangle = \frac{\sum_i V_i \rho_{i}}{\sum_i V_i}
\end{equation}

\begin{equation} \label{eqn:delta}
    \Delta \equiv \Delta_{\mathrm{V}} = \frac{\rho_i}{\langle \rho \rangle}
\end{equation}






We use this overdensity calculation to create 2D histograms of the fraction of neutral hydrogen for each particle in the simulation as a function of overdensity for four redshift snapshots, as shown in Figure \ref{fig:ezra_plot}. 
The color bar represents the number of particles in that bin. The gas is mostly neutral at high redshift, with reionization (transition of particles from top to bottom in this figure) working from lower to high density regions (left to right). This supports the outside-in reionization model where low-density regions ionize first, and high-density regions remain neutral longer, even after the volume average universe is mostly ionized ($z=6$).

For a simple quantitative metric of the reionization topology, we compute the ratio of the mass-averaged to volume-averaged neutral fraction over redshift for our simulation, shown in Figure \ref{fig:topology}. When this ratio is less than one (orange shaded region), this indicates the mass-averaged neutral fraction is decreasing faster, implying that dense regions are undergoing reionization faster at these times (inside-out reionization). On the other hand, a ratio greater than one (green shaded region) indicates that the volume-averaged mass fraction changes faster and therefore implies that the least dense regions are reionizing faster (outside-in reionization). From about $16<z<11$, the ratio is slightly below unity, indicating an early inside-out phase of reionization. When $z<11$, this ratio quickly increases, demonstrating the dominant effect of outside-in reionization.

Initially, radiation escapes from ionizing sources and ionizes the immediate, dense surroundings. The radiation then leaks directly into the least dense regions, and reionization occurs quickly in these regions due to the smaller number of particles to ionize and the low recombination rate. Finally,
reionization progresses towards higher density regions, shown in Figure \ref{fig:ezra_plot} as a yellow diagonal feature. This demonstrates that neutral fraction increases with density. This is consistent with particles being in an ionization equilibrium with a uniform ultraviolet background (UVB) as higher density gas has a high recombination rate.

Our results agree with previous findings from \cite{finlator_late_2009}, who find an IOM reionization topology. In this model, reionization initially starts in the dense regions immediately surrounding ionizing sources, then leaks into the least dense regions, and finally progresses from the least dense to intermediate density regions of the universe, as the lower density regions ionize faster due to having fewer atoms to ionize per volume as well as a lower recombination rate. This process creates ionized bubbles, or cosmological HII regions, which eventually overlap and cause ionization fronts to move back in towards denser gas. As a result, small high-density clouds of neutral gas can remain owing to self shielding. The initial inside-out phase is short-lived, and the more dominant effect is from the outside-in phase of reionization.

\cite{doughty_evolution_2019} demonstrate that for outside-in reionization topology, the CGM will remain neutral even after the IGM is largely ionized. This implies that the discovery of a significantly neutral CGM does not guarantee that reionization is incomplete, but that this may be self shielded neutral gas clouds remaining after reionization. Outside-in topology also predicts weak relations between galaxies and the ionization state of their associated CGM, as the CGM is likely ionized by the large scale ultraviolet background, not just the local galaxy \citep{doughty_evolution_2019}. However, the topology in our simulation is not purely this simple, as there is ionized overdense gas at $z=10$ and neutral overdense gas at $z=5.5$, suggesting the need for more careful analysis and study of individual galaxies.

\subsection{Halos and CGM Characteristics}\label{subsec:cgm}

The nontrivial reionization topology shown in the previous analysis compels us to adopt a halo-centric perspective in order to compare the CGM to IGM evolution. 
We first identify galaxies using a 5th-order spline kernel interpolative denmax (SKID, \cite{governato_local_1997}). Then, using a spherical overdensity calculation, we identify each galaxy's host dark matter halo \citep{finlator_faint_2020}. We have found that the resulting dark matter halo mass function is in excellent agreement with the Sheth-Tormen mass function, indicating that the resulting halo-galaxy association is reasonable. 

This simulation is limited by the dark matter halo mass function resolution, the galaxy resolution, and the HI cooling mass threshold. Therefore, to select only resolved halos, we include only galaxies with halo masses larger than $10^{8.5} M_\odot$, the largest of the previously mentioned limitations, at redshifts of 10, 8, 6, and 5.5.

\begin{figure}
\centering{\includegraphics[width=0.495\columnwidth]{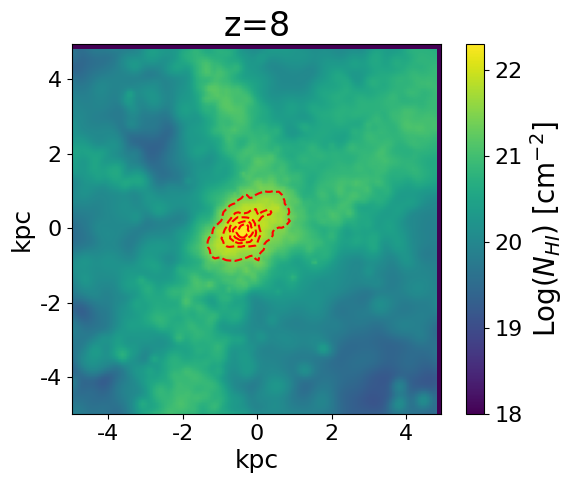}}
\centering{\includegraphics[width=0.495\columnwidth]{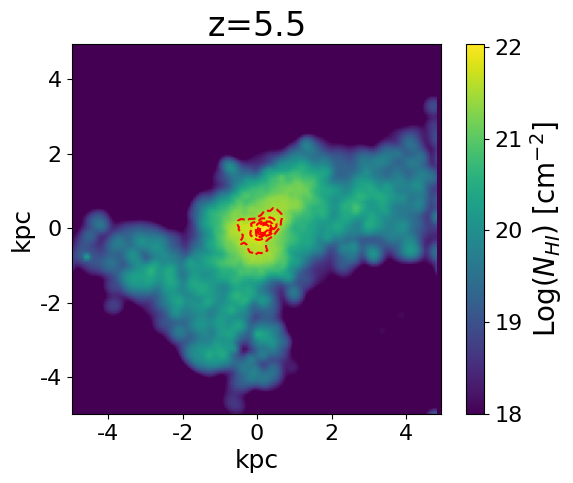}}
\caption{Visual representations of column density of neutral hydrogen for $z=8$ and $z=5.5$ massive halos. Each have a halo mass of $M \approx 3.3*10^{10} M_{\odot}$. The red contours correspond to regions with high star formation rates. Higher column densities of neutral hydrogen appear around the $z=8$ halo than around the $z=5.5$ halo while the masses remain similar, implying that gas around the $z=5.5$ has a lower neutral fraction as a result of reionization. 
\label{fig:halos}}
\end{figure}

We begin our analysis by computing column density of neutral hydrogen in the foreground of \td\, simulated galaxies, as this is a key parameter when comparing to observations. To do this, we first isolate a cubical region around each halo, with side lengths corresponding to 2 times the virial radius of the halo in order to capture the CGM region. We then divide each cube into $128^3$ cells, and, using our simulation  outputs, we compute the amount of gas, amount of neutral hydrogen ($\frac{N_{HI}}{volume}$), and the star formation rate in each cell. We collapse these cubes in one dimension to create 2D plane figures, which allows us to create the visual representations of our halos as shown in Figure \ref{fig:halos}. This results in values for column density of neutral hydrogen ($N_{HI}$ [$cm^{-2}$]) and star formation rate ($\sigma_{SFR}$) for each $x$ and $y$ position in our 2D halo image.


To make a more direct comparison of our computed column densities to inferences from observations (e.g. \cite{heintz_strong_2024}), we weight our simulated column densities by the star formation rate following Equations \ref{eqn:NHI_unsimp} and \ref{eqn:NHI}. Observationally, the source of light for detecting these PDLAs is the galaxy itself, from the stars within the galaxy. Therefore, only gas in the foreground of regions with stars can be seen. We account for this in our simulation by weighting by the star formation rate in each cell, as bright, young stars are known to exist in these areas. 

\begin{equation}\label{eqn:NHI_unsimp}
   \langle N_{HI} \rangle = \frac{\frac{1}{2} \int{N_{HI}(x,y) \sigma_{SFR}(x,y) dxdy}}{\int{\sigma_{SFR}(x,y) dxdy}}
\end{equation}

We can simplify Equation \ref{eqn:NHI_unsimp}, noting that in our simulation $dxdy$ is always just $dA$, allowing us to express this as a sum rather than an integral. We also include a factor of $\frac{1}{2}$ 
to average the foreground and background column density for each galaxy. This, combined with the large number of galaxies included in this study, should account for variations in the population's geometry. \cite{gelli_neutral_2025} finds that this is an important factor when computing the column densities, as sightline-to-sightline variations due to complex IGM geometry can change the column density measurement by up to about 1.5 dex.


\begin{equation}\label{eqn:NHI}
       \langle N_{HI} \rangle = \frac{\frac{1}{2} \sum N_{HI}(x,y) \sigma_{SFR}(x,y) dA} {\sum\sigma_{SFR}(x,y) dA}
\end{equation}


For the remaining properties, we need to access information from individual particles within halos, and thus analyze the snapshot outputs. Using the halo catalog outputs to find the central coordinates and virial radius of each halo, we collect information from all gas, dark matter, and star particles within a spherical volume (using the virial radius) of each halo central position. Periodic boundaries were not imposed, which results in halos that lie near simulation boundaries being discarded. This reduces the computation time substantially while excluding less than half a percent of our selected halos.

We compute the ratio of neutral hydrogen to total hydrogen weighted by particle mass, or mass weighted neutral fraction ($\langle f_{HI} \rangle$, Equation \ref{eqn:NF}), as well as the ratio of the gas mass to the total mass of each halo ($f_g$, Equation \ref{eqn:GF}). 

\begin{equation}\label{eqn:NF}
   \langle f_{HI} \rangle = \frac{\sum_i m_i (\frac{n_{HI}}{n_H})_i}{\sum_i m_i}
\end{equation}

\begin{equation}\label{eqn:GF}
   f_g = \frac{\sum m_{g,i}}{\sum m_{g,i} + \sum m_{dm,i} + \sum m_{*,i}
   }
\end{equation}

The predicted neutral hydrogen column density, halo mass, stellar mass, and redshift are also found for each halo. 
We also use these parameters to compare with motivating JWST observations.

\begin{figure}
\centering{\includegraphics[width=0.998\columnwidth]{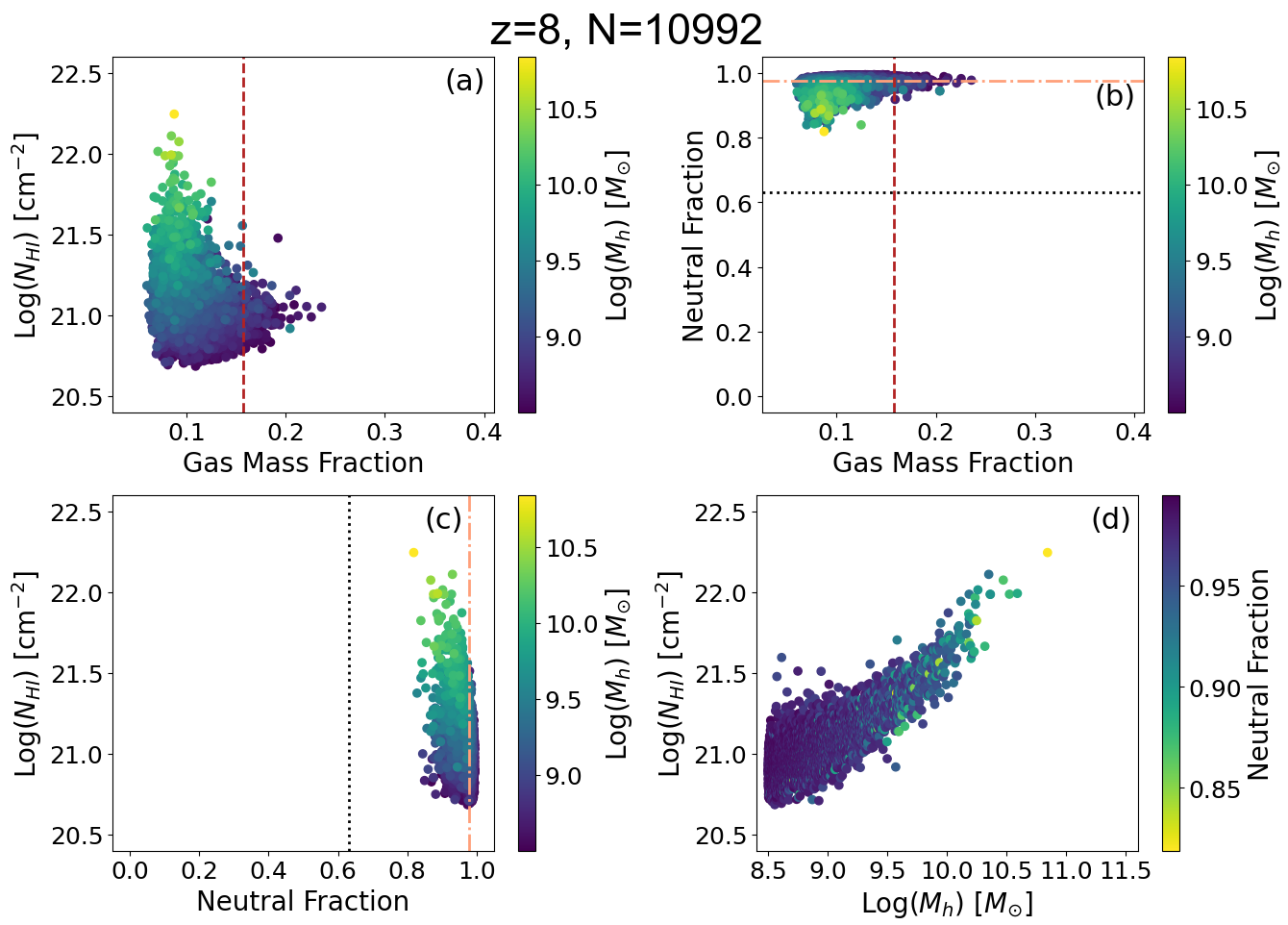}}
\caption{Relationships between the gas mass fraction, column density of neutral hydrogen, mass weighted local neutral fraction, and halo mass for $z=8$ halos. The dashed red line is the cosmic baryon fraction for this simulation ($\Omega_b / \Omega_m \approx 0.1573$), the dash-dotted orange line is the averaged local neutral fraction, and the dotted black line is the volume-averaged neutral fraction at $z=8$. 
\label{fig:snap_analysis_z8}}
\end{figure}

For a visual impression of how the neutral CGM recedes through the latter stages of reionization, we show images of similarly massed halos at two different redshifts, $z=8$ and $z=5.5$, shown in Figure \ref{fig:halos}. The plots are scaled so that each side is twice the virial radius of the central halo. Comparing the two, the CGM around the $z=8$ halo has a higher column density of neutral hydrogen that extends farther out around the galaxy than the $z=5.5$ galaxy does. This may be an indicator that the $z=8$ galaxy's environment is not yet reionized, whereas the $z=5.5$ galaxy's environment is actively undergoing reionization. 
The star-forming regions of each galaxy are shown in red contours. 

\begin{figure}
\centering{\includegraphics[width=0.998\columnwidth]{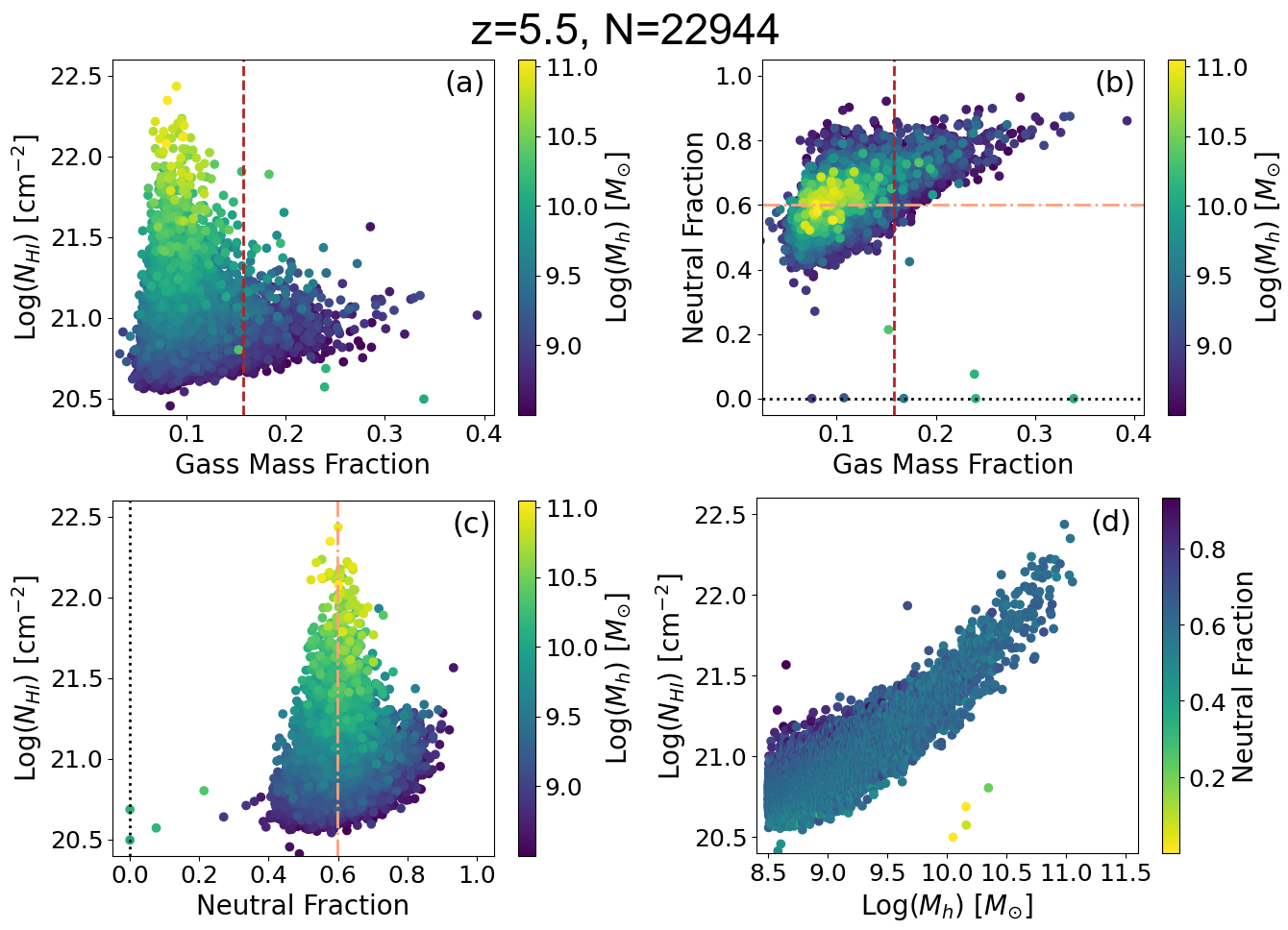}}
\caption{Similar to Figure \ref{fig:snap_analysis_z8} for $z=5.5$ halos. 
\label{fig:snap_analysis_z5_5}}
\end{figure}

To investigate whether high HI columns reflect high gas fractions, high local neutral fractions, or the presence of a neutral large-scale environment, 
we use the outputs of our individual halo analysis to study relationships between the properties listed above for our sample of galaxies at $z=8$ and $z=5.5$, as shown in Figures  \ref{fig:snap_analysis_z8} and \ref{fig:snap_analysis_z5_5}. The dashed black line is the cosmic baryon fraction, $\Omega_b / \Omega_m \approx 0.1573$ for this simulation, which would be the gas fraction if baryons were associated 1-to-1 with dark matter particles. The brown dash-dotted line is the average local neutral fraction, and the dashed gray line is the volume averaged (global) neutral fraction for the respective redshifts.

Panel (a) in the upper left of Figure \ref{fig:snap_analysis_z8} shows no obvious relationship between gas mass fraction and the neutral hydrogen column density (\NHI). On the contrary, the halos with the highest \NHI  actually have sub-cosmic baryon (gas) fraction. 
This conflicts with the hypothesis that the observed PDLAs simply trace systems with more gas. A glance at the color bar suggests that, instead, high \NHI appears more directly associated with high halo masses.

The result that high \NHI is associated with low gas fractions suggests that large HI columns may reflect a high neutral fraction. In order to test this possibility, we show in panel (b), the top-right of Figure \ref{fig:snap_analysis_z8},
that low neutral fractions are associated with low gas fractions and high halo masses. This is consistent with a scenario where halo environments have already processed more gas into stars, creating more ionizing flux, and therefore ionizing more of their CGM. This is most apparent for the more massive halos, while the low-mass halos show more scatter. This may be due to low-mass halos occupying a broad range of environments and formation histories, whereas high-mass halos only occupy overdense environments. 
Broadly, however, the CGM remains almost entirely neutral prior to the completion of reionization.

A more direct way to test whether large HI columns reflect high neutral fractions is to compare them directly, as in panel (c) in the bottom left of Figure \ref{fig:snap_analysis_z8}. This shows no trend between the neutral fraction and \NHI, so we can conclude that high column densities of neutral hydrogen do not simply reflect higher neutral fractions. 

Having demonstrated that \NHI does not predominantly reflect the gas fraction or the local neutral fraction, we ask whether it simply traces overall halo mass. Panel (d) in the bottom right shows a clear correlation between halo 
mass and \NHI. There is a slight gradient in the color, denoting neutral fraction, tracing the slight trend found in the top right panel as well. In all, the neutral hydrogen column density is not driven by gas mass fraction or local neutral fraction. Instead, the host halo mass has the most clear correlation with \NHI, therefore the most massive halos will have the highest column density of neutral hydrogen despite having a low neutral fractions and low gas fractions.  


Figure \ref{fig:snap_analysis_z5_5} repeats the above analysis post-reionization. A noticeable similarity between Figures \ref{fig:snap_analysis_z8} and \ref{fig:snap_analysis_z5_5} is that in both, panel (a) shows no clear relation between the gas mass fraction and \NHI. Again, this suggests that we are not simply seeing larger column densities due to larger volumes of gas.

\begin{figure}
\centering{\includegraphics[width=0.995\columnwidth]{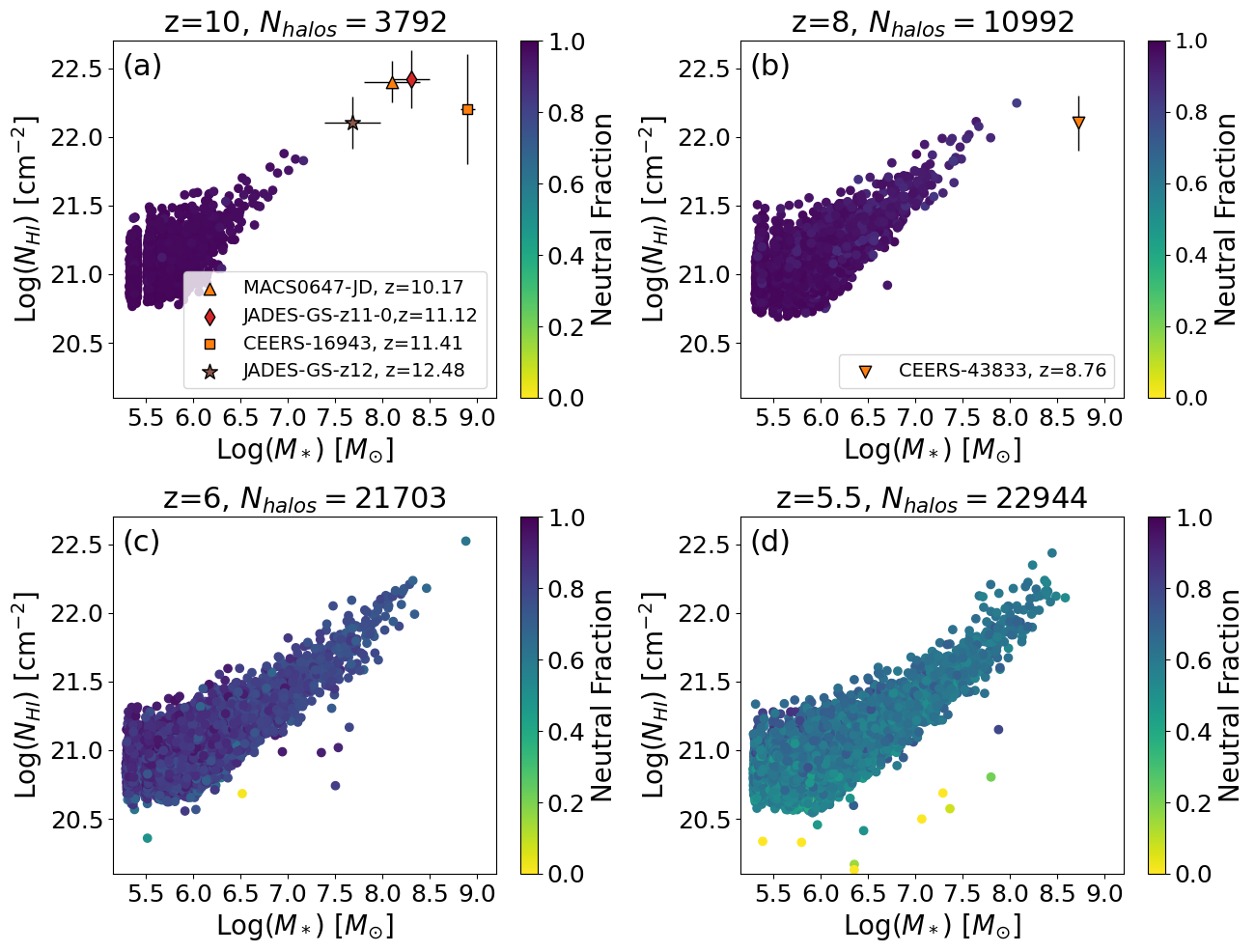}}
\caption{Grid showing column density of neutral hydrogen as a function of stellar mass for halos in various redshift bins. Observations from \cite{heintz_strong_2024} (orange), \cite{hainline_searching_2024} (red), and \citep{deugenio_jades_2024} (brown) are overlaid with error bars.  
\label{fig:observables}}
\end{figure}

We compare properties of halos during reionization ($z=8$) and after reionization is complete ($z=5.5$) in panels (b) and (c) of Figures \ref{fig:snap_analysis_z8} and \ref{fig:snap_analysis_z5_5}. In panel (b) of Figure \ref{fig:snap_analysis_z5_5}, there is an emerging correlation between gas mass fraction and neutral fraction. If the gas fraction is low, the total hydrogen density is low and the recombination rate is low, therefore ionization equilibrium shifts towards a low neutral fraction. Therefore, this correlation is likely a consequence of ionization equilibrium with a weak background. 
Panel (c) of Figure \ref{fig:snap_analysis_z5_5} shows no strong trend between neutral fraction and \NHI, but compared to the same panel of Figure \ref{fig:snap_analysis_z8}, the distribution has moved to the left, indicating that these halos are generally more ionized. This is also visible in panel (b) as the distribution moving down. In panels (b) and (c) of Figure \ref{fig:snap_analysis_z5_5} the spread of points is wider than in the corresponding panels of Figure \ref{fig:snap_analysis_z8}, demonstrating that reionization is a non-uniform process.

Finally, panel (d) of Figure \ref{fig:snap_analysis_z5_5} shows the same clear correlation between mass and \NHI as Figure \ref{fig:snap_analysis_z8}, but this time there is no clear relation to the local neutral fraction. Again, halo mass is the best predictor of \NHI.  

In order to suggest how the simulated evolution may be probed observationally, we compare \NHI with galaxy stellar mass at four different redshifts: $z=10, 8, 6 \text{ and } 5.5$ (Figure \ref{fig:observables}). There is a clear correlation between stellar mass and \NHI much like in the bottom right panels of Figures \ref{fig:snap_analysis_z8} and \ref{fig:snap_analysis_z5_5}. 
Neutral fraction is also shown on the colorbar for each panel, demonstrating a slow transition from mostly neutral halos at $z=10,8$ to partially ionized halos at $z=5.5$. 
Observed galaxies from \cite{heintz_strong_2024}, \cite{hainline_searching_2024}, and \cite{deugenio_jades_2024} have been added, with observational error bars shown. Galaxies with a wide range of observed redshift values are included in the $z=10$ panel to provide as many observations as possible. The physical time difference from $z=10$ to $z=8$ is comparable to that of $z=10$ to $z=13$, so we believe including the $z=12.5$ galaxy from \cite{deugenio_jades_2024} is not unwarranted for the $z=10$ panel, given the limited data. 



Generally there is a shift toward lower column densities of neutral hydrogen at lower redshifts, which is evidence of reionization, but this trend has less of an overall effect than the dominant trend that column density depends most on mass.

\subsection{Inferring the Average Neutral Fraction and Observational Feasibility}\label{subsec:fit}

To quantify the relationships between the column density of neutral hydrogen, halo or stellar mass, and the volume averaged neutral fraction of hydrogen, we fit our data to the following functional form, where $N_{HI}$ is the column density of neutral hydrogen in units of per centimeter squared ($cm^{-2}$), $M$ is the mass in units of solar masses ($M_{\odot}$) and $<x_{HI}>$ is the average neutral fraction (volume-averaged in this case). 

\begin{equation}\label{eqn:fit}
    \log_{10}(N_{HI}) = a + b \,  \log_{10}(M) + c <x_{HI}>
\end{equation}

Using a least squares method and halo mass, we find best fit values for our model to be: $a = 17.283\pm0.010$, $b=0.407\pm0.001$, $c=0.212\pm0.001$. Replacing halo mass with stellar mass, we find similar parameter values of: $a = 19.515\pm0.007$, $b=0.242\pm0.001$, $c=0.211\pm0.002$. Nonzero values for b and c show a clear correlation between these parameters, with halo or stellar mass being the more dominant effect. However, because c is still nonzero, the volume average neutral fraction is clearly related to the column density of neutral hydrogen. This means that \NHI, as inferred from DLA strength, can be used to trace the progress of reionization.

This relation is plotted in Figure \ref{fig:function} with the redshift and corresponding average neutral fraction for three different snapshots in our simulation. We plot three lines corresponding to characteristic halo masses of $M_h = 10^9, 10^{9.5}$, and $10^{10} M_{\odot}$. We then overplot a sample of galaxies with masses similar to these characteristic masses and their corresponding column densities, to show the scatter in our relationship. This scatter may include various astrophysical effects, such as halo environment, the geometry of our sightlines, and galaxy-to-galaxy variations within our simulation. This scatter will make this relationship more difficult to use in tracing reionization observationally. 



To test how feasible this method may be for observers, we perform a simplistic propagation of errors calculation to find an estimated number of observed galaxies needed to measure $\langle x_{HI} \rangle$ to a given fractional error. First, we rearrange Equation \ref{eqn:fit} to isolate the neutral fraction for each galaxy based on its column density and mass: 

\begin{equation}
    x_{HI} = \frac{\log_{10}(N_{HI}) - a - b \, \log_{10}(M_*) - \epsilon}{c}
\end{equation}

Where $\epsilon$ is a term that depends on the scatter of our simulated galaxies. Ignoring covariances, we find that a propagation of errors yields the following expression:

\begin{multline}\label{eqn:prop_errs}
    \sigma^2_{x_{HI}}(z) = \frac{1}{c^2}[ \sigma^2_{\log(N_{HI})} +  b^2 \sigma^2_{\log(M_*)}(z) + \sigma^2_s (z) + \sigma^2_a + \\ (\langle \log(M_*) \rangle (z))^2 \sigma^2_b + (\langle x_{HI} \rangle (z))^2\sigma^2_c]
\end{multline}

Where we now have $\sigma_s$, which is the scatter term. We can measure this by finding the residuals ($\Delta$, Equation \ref{eqn:residuals}) of our fit (Equation \ref{eqn:fit}) and taking the standard deviation of those residuals.

\begin{equation}\label{eqn:residuals}
    \Delta(z) = \log(N_{HI})_{true} - \log(N_{HI})_{model}
\end{equation}


For redshifts of 10, 8, and 6, we find the mean residuals are $0.010$, $-0.005$, and $0.031$, and $\sigma_s$ values are $0.12351$, $0.12252$, and  $0.12251$ dex. This captures our galaxy-to-galaxy variations, but this is likely an underestimate of the true astrophysical scatter. For example, \cite{gelli_neutral_2025} find a 0.5-1.5 dex scatter due to sightline variations for each galaxy, which is not accounted for here. This may also not fully account for additional scatter from cosmic variance, as \td \, has a limited box size and cannot produce the most massive halos. Therefore, going forward, we note that this exercise will result in a lower bound for the number of galaxies needed in an observational study.


{We calculate $\Delta$ (and therefore $\sigma_s$), $\langle \log(M_*) \rangle$ (for resolved galaxies), and $\langle x_{HI} \rangle$ for each of our redshift snapshots, and have $\sigma_a$, $\sigma_b$, and $\sigma_c$ from the errors on the fit parameters. We assume that $N_{HI}$ can be measured to high precision, so we can disregard this term. We take $\sigma_{\log(M_*)}$ to be 0.3 dex, corresponding to the maximum scatter in residuals in stellar mass for all SFH parameterizations and redshifts ($z=5-10$) from \cite{cochrane_high-z_2025}.


Using the resulting estimate of the error on $x_{HI}$, we can determine the number of galaxies needed to compute $\langle x_{HI} \rangle$ to some fractional error ($f$), following:

\begin{equation}
   \sigma_{\langle x_{HI} \rangle} (z) = \frac{\sigma_{ x_{HI}}(z)}{\sqrt{N}} 
\end{equation}

\begin{equation}
   f = \frac{\sigma_{\langle x_{HI} \rangle} (z)}{\langle x_{HI} \rangle(z)} = \frac{\sigma_{ x_{HI}}(z)}{\langle x_{HI} \rangle(z) \sqrt{N}} 
\end{equation}

\begin{equation}
   N(z) = (\frac{\sigma_{ x_{HI}}(z)}{\langle x_{HI} \rangle (z)f})^2
\end{equation}

\begin{figure}
\centering{\includegraphics[width=0.998\columnwidth]{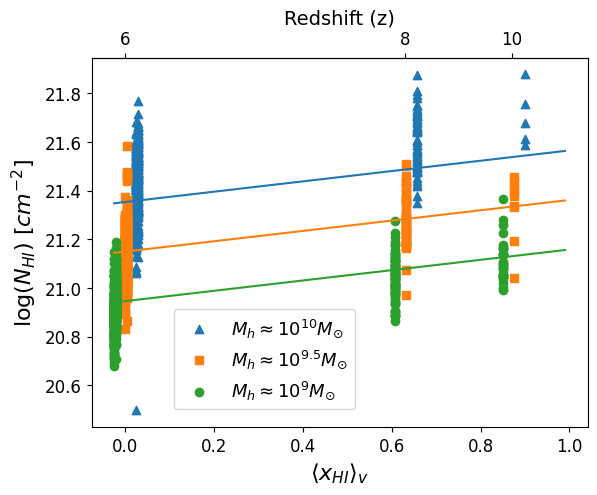}}
\caption{Relations between column density of neutral hydrogen (log$(N_{HI})$), halo mass ($M_h$), and volume averaged neutral fraction ($\langle x_{HI} \rangle_v$, which is related to redshift or time evolution). Halo mass is the dominant effect, shown as differences in the y-intercept for different mass bins. Neutral fraction is also related to column density, showing a clear trend with time as reionization progresses. Note that there is a slight offset added to the the neutral fraction values for clarity. 
\label{fig:function}}
\end{figure}

The resulting required number of observed galaxies (N) is shown in Figure \ref{fig:obs_feasibility} for different choices of fractional precision on $\langle x_{HI} \rangle$. We base our feasibility ranges on current and projected observations, with the current survey number ($N\approx25$) from the analysis of \cite{pollock_characterising_2026}, which finds 27 high-confidence galaxy PDLAs. We estimate the JWST-era feasibility ($N\approx150$) from noting that there are approximately 30 known galaxy PDLAs currently (after 4 years of JWST), and projecting this forward assuming JWST continues detecting high redshift galaxies for the next 10-15 years. The future mission feasibility is more qualitative, but a goal of $1000$ observed galaxy PDLAs seems reasonable for a targeted future mission. Overall, we find that this method appears to be promising for higher redshifts where $\langle x_{HI} \rangle$ is relatively large. This method, however, quickly becomes less sensitive when the neutral fraction is low, due to $\langle x_{HI} \rangle^{-2}$ scaling with N. Therefore, we suggest that this new method of tracing reionization using galaxy PDLAs may be well suited to constraining early reionization.

However, these findings do have some caveats. The estimated numbers of observations required are effectively lower limits, as we do not explicitly account for additional scatter in this relation due to sightline-to-sightline variations, environmental differences, or uncertainties in measuring the column density of neutral hydrogen. We also do not account for any covariances between sources of error. This may cause significant additional scatter, which could affect our estimated number of required observations by an order of magnitude or more. 



\begin{figure}
\centering{\includegraphics[width=0.998\columnwidth]{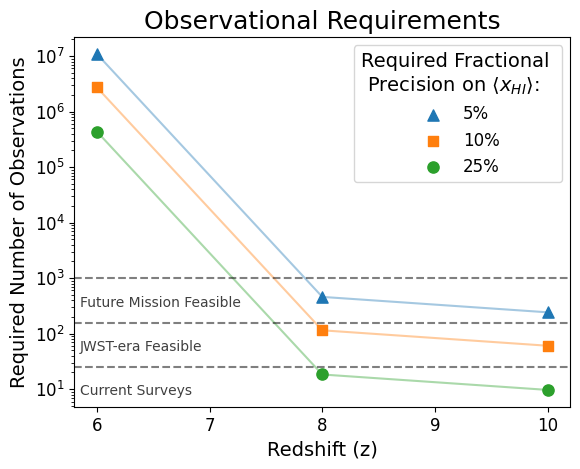}}
\caption{The required number of observations as a function of redshift and required fractional precision ($f$) on $\langle x_{HI} \rangle$. The three curves correspond to $f=0.05, 0.10,$ and $0.25$. Horizontal bands represent approximate observational regimes: current surveys ($N\approx25$), JWST-era feasible ($N\approx150$), and potential future missions ($N\approx1,000$). The steep increase in required number of observations at low redshift is due to the low neutral fraction as reionization is nearly complete, as $N \propto \langle x_{HI} \rangle^{-2}.$ The required number of observations are lower bounds, as we do not account for signtline-to-sightline variations (shown to be significant in \cite{gelli_neutral_2025}), have limited dynamic range and cosmic variance, and assume no covariances between parameters.
\label{fig:obs_feasibility}}
\end{figure}




\section{Discussion} \label{sec:discussion}

From our snapshot analysis, we find a clear relation between the column density of neutral hydrogen and the halo (or stellar) mass. Conversely, we find that the gas fraction is not strongly correlated with the column density of neutral hydrogen and has a wider range at lower redshift. This shows that the ratio of gas to stars and dark matter is not the main factor contributing to the high column densities in the simulation. We also find that local neutral fraction has a slight anti-correlation with mass at high redshift ($z=10$ and $z=8$), but this is not present at low redshift ($z \leq 6$). However, the global neutral fraction can be estimated from the column density of neutral hydrogen, showing a clear tie between galaxy PDLAs and HI reionization. 

While the simulated galaxies do not overlap in stellar mass with the observed systems, the observed systems' foreground \NHI appears consistent with an extrapolation of the simulated trend. There is also a general agreement with predicted column densities of neutral hydrogen from \cite{umeda_jwst_2024}, which shows galaxies from $7 < z < 12$ with column densities from $18 <$ log(\NHI) $[cm^{-2}] < 23$ and increasing column density as redshift increases. We also see a broad agreement with observational work from \cite{pollock_characterising_2026}, who find that the average column density of neutral hydrogen increases with redshift, with $N_{HI} = 10^{21.71}, 10^{22.24},$ and $10^{22.37} cm^{-2}$ for redshift bins $z=9-10, \, z=10-12,$ and $z>12$. \cite{umeda_jwst_2024} also notes that \NHI is unique to individual galaxies, agreeing with our analysis that \NHI and therefore DLAs are most strongly associated with individual galaxy characteristics (halo or stellar mass).

We also find a dependence of $N_{HI}$ on $M_h$ similar to that in \cite{gelli_neutral_2025}, who investigate halos with much larger masses ($\log(M_h) \approx 9.5-11.5$). In the form $\log(N_{HI}) = b \, \log(M_h)$, \cite{gelli_neutral_2025} find $b=0.38$, where we find $b=0.407$ (see Section \ref{subsec:fit} for details on this calculation). We find this agreement over many orders of magnitude in mass to be very reassuring, and this may even be related to the scaling of halo mass with the virial radius as discussed in \cite{gelli_neutral_2025}. 


Although we are encouraged by the agreement between our results and literature results on PDLAs \citep{gelli_neutral_2025, umeda_jwst_2024}, there are still some caveats to this work. This includes variations due to the geometry of our sightlines, which may not be fully accounted for in our models. Recent work from \cite{gelli_neutral_2025} has found that sightline-to-sightline differences from a more complex ISM geometry may contribute up to 1.5 dex variations in the measured column density of neutral hydrogen. In this work, we do not quantify sightline-to-sightline variations, but instead look at a large number of galaxies to account for different sightlines, which may not capture the full scatter, as we have less resolving power in the ISM compared to zoom simulations. The large-scale environment may also be leading to significant scatter, and observational efforts to characterize the impact of overdensities on PDLA strength have been performed by \cite{terp_all_2026}, finding that some overdensities include many members that display these PDLAs. This simulation, however, has a limited box size of 16.5 cMpc/h side lengths and does not produce these very overdense environments, which could imply that we are still underproducing expected scatter, especially at the high mass end.


Recent literature studies using simulations to study galaxy PDLAs offer a wide perspective on the importance of studying these objects. \cite{umeda_probing_2025} argue that these PDLAs are not necessary to explain the observed spectra, and that the damping from the neutral hydrogen in the IGM alone is sufficient, given a more accurate inhomogeneous IGM. On the other hand, \cite{keating_jwst_2024} show that the modeled damping wings still require a PDLA to match the strength of damping in the observations. \cite{huberty_pitfalls_2025} demonstrates the complex nature of deriving the column density of neutral hydrogen for local gas and the neutral fraction of the foreground IGM as these quantities are highly degenerate. This degeneracy further complicates the use of galaxy PDLAs in tracing cosmic HI reionization. 

Observers are also working to build a larger sample of galaxy PDLAs using JWST (e.g. \cite{ pollock_characterising_2026, terp_all_2026}).  Interestingly, there has been little evidence of a relation between the column density of neutral hydrogen and the UV magnitude (often used to infer halo mass) from \cite{pollock_characterising_2026} and \cite{terp_all_2026}, as opposed to this work and \cite{gelli_neutral_2025}, where this is the dominant trend. Additionally, there are speculations on the physical origins of this neutral gas different to the CGM origin assumed here, including cold neutral filaments penetrating overdensities \citep{terp_all_2026}, as well as discussion regarding how only some galaxies seem to display these PDLAs and not others, including the potential presence of ionized bubbles or contamination from unresolved Lyman-$\alpha$ emission \citep{pollock_characterising_2026}.  



Here, we provide some ideas as to what physical conditions may be implied by the relation between these galaxy PDLAs, redshift, and halo mass within our simulation. Because trends change with redshift, we break this into high redshift (before/ongoing reionization) and low redshift (after reionization). At high redshifts, low-mass halos are found everywhere, but massive halos require overdense environments. This may be the cause of higher scatter in gas mass fraction and neutral fraction for lower mass halos. For higher mass halos, the overdense environment implies more local ionizing sources and therefore a stronger local ionizing background. This leads to an overall lower neutral fraction. However, because these halos contain substantially more gas, even with a low neutral fraction, there is sufficient neutral gas to produce higher column densities of neutral hydrogen. At lower redshifts, the ionizing background is more uniform and more intense on average, so both high-mass and low-mass halos will be surrounded by ionizing radiation. We again predict more total gas and therefore more total neutral gas for more massive halos, so we continue to expect that more massive halos will have higher column densities of neutral hydrogen.


As for the topology of reionization, our results 
allign with \cite{finlator_late_2009}, which finds an IOM reionization. 
This is characterized by a brief period of inside-out reionization, followed by a more dominant and long-lasting outside-in phase. In this topology, ionizing sources located within dense media emit photons which first ionize the local surroundings (the short inside-out phase) enough for some ionizing photons to escape to the IGM. Then, reionization follows a more outside in topology, where these escaping photons are able to ionize the low density IGM relatively quickly, as low density regions have fewer atoms to ionize per volume and a lower recombination rate \citep{doughty_evolution_2019}. This process creates cosmological HII regions in the IGM, which increase in size along the direction of lowest density \citep{miralda-escude_reionization_2000}. Eventually these regions begin to overlap, and the ionization fronts move to regions of higher densities. Finally, the IGM and CGM are reionized, leaving only a few high density clouds that can remain neutral due to self-shielding.

\section{Summary and Future Work} \label{sec: summary}

The key takeaways from this project are:
\begin{enumerate}
    \item This simulation predicts an inside-out-middle reionization topology.
    \item In the immediate aftermath of reionization ($z=5.5$), the CGM remains significantly neutral, and in ionization equilibrium with the UVB, such that halos with higher gas mass fractions are locally more neutral.
    \item Our predictions do not support the interpretation that high column densities of neutral hydrogen are associated with high local gas mass fractions or high local neutral fractions.
    \item We find a clear relation between the column density of neutral hydrogen [$cm^{-2}$], halo (stellar) mass [$M_{\odot}$], and volume-averaged neutral fraction. We quantify this using a functional form: $\log_{10}(N_{HI}) = a + b \,  \log_{10}(M) + c <x_{HI}>$ and find best fit values:  $a = 17.283\pm0.010 \, (19.515\pm0.007)$, $b=0.407\pm0.001 \, (0.242\pm0.001) $, $c=0.212\pm0.001 \, (0.211\pm0.002)$. 
    \item Given precise estimates of halo or stellar mass, galaxy PDLAs provide a new method for tracing the progress of reionization.
    \item We believe that this method is observationally feasible with ongoing JWST observations, particularly for constraining early reionization when the average neutral fraction of the universe is high.
\end{enumerate}


In the future, it would be beneficial to focus on larger box sizes so that we can account for larger halos in our simulations. This will allow us to compare to JWST observations where the galaxies are significantly larger than those in this simulation. It appears that the observations fall on an extended trend line from our work (Figure \ref{fig:observables}), which we find is in agreement with simulations featuring larger halos \citep{gelli_neutral_2025}, but this should certainly be tested. Larger halo masses would also allow us to test if the relation between $\log_{10}(N_{HI})$ and $\log_{10}(M_h)$ deviates from linear at the high mass end. It is most important to obtain simulations with larger halos at high redshifts, so that we can explore the following questions: 
Are simulations under producing large halos? Are observed halos significantly bigger than expected? Are the large observed halos just an observational bias, as larger galaxies are brighter and therefore easier to observe? We could also further compare this work to other simulations to determine which inputs may be causing differences in reionization topology. 
Finally, can we compare our predicted reionization topology to observations, and how can these comparisons inform our models?

\section{Acknowledgments}

This work used resources from the New Mexico State University High Performance Computing Group, which is directly supported by the National Science Foundation (OAC-2019000), the Student Technology Advisory Committee, and New Mexico State University and benefits from inclusion in various grants (DoD ARO-W911NF1810454; NSF EPSCoR OIA-1757207; Partnership for the Advancement of Cancer Research, supported in part by NCI grants U54 CA132383 (NMSU)). KF and MS gratefully acknowledge support from program \#JWST-AR-05779.001-A, support for which was, in turn, provided by NASA through a grant from the Space Telescope Science Institute, which is operated by the Association of Universities for Research in Astronomy, Inc., under NASA contract NAS 5-03127. 

MS and KF also thank Kasper E. Heintz for insightful discussions that helped motivate this work. MS also thanks the attendees of the Charting Cosmic Dawn in Copenhagen 2026 meeting for productive conversations. In addition, the authors thank Len Hillhouse for providing feedback and ideas during group research meetings. Finally, we also thank the referee for the detailed and informative report.



This research has been enabled by the NASA Astrophysics Data System and the arXiv eprint service. The Cosmic Dawn Center is funded by the Danish National Research Foundation under grant DNRF140.

\section{Author Contributions}

\textbf{MS}: led the analysis and writing of the paper; \textbf{KF}: provided support to analysis, reviewed and edited the text, oversaw the project progress, and acquired the funding to support the project; \textbf{SK \& EH}: provided support to analysis, and reviewed and edited the text.

\bibliography{reionization_new}{}
\bibliographystyle{aasjournal}

\end{document}